\documentclass[aps,nofootinbib,superscriptaddress, showpacs,preprintnumbers, nofootinbibt,twocolumn]{revtex4-2}

\usepackage{epsfig}
\usepackage{multirow}
\usepackage{subcaption}
\usepackage{eurosym}
\usepackage{dcolumn}
\usepackage{bm}
\usepackage{enumerate}
\usepackage{float}
\usepackage{epstopdf}
\usepackage{amsmath}
\usepackage{bm}
\usepackage{amsfonts}
\usepackage{amssymb}
\usepackage{graphicx}
\usepackage{alphalph,mathtools}
\usepackage{etoolbox}
\usepackage{color}
\usepackage{booktabs}
\usepackage{hyperref}
\hypersetup{colorlinks,citecolor=blue}
\usepackage{footnote}
\usepackage{makecell,tabularx}

\usepackage{graphicx} 

\def\be{\begin{equation}}
\def\ee{\end{equation}}
\def\bea{\begin{eqnarray}}
\def\eea{\end{eqnarray}}

\begin{document}

\title{The New Black Hole Solution with Anisotropic Fluid in \( f(R, \mathcal{L}_m, T) \) Gravity: Thermodynamics } 

\author{Aniruddha Ghosh}
 \email{ruddha.g@gmail.com}
 \affiliation{%
 Department of Mathematics, Indian Institute of Engineering Science and Technology, Shibpur, Howrah-711103, India.\\ 
}%
\author{Ujjal Debnath}%
 \email{ujjaldebnath@gmail.com}
\affiliation{%
 Department of Mathematics, Indian Institute of Engineering Science and Technology, Shibpur, Howrah-711103, India.\\ 
}%

\begin{abstract}
In this work, we derive a new class of analytic black hole solutions within the framework of \( f(R, \mathcal{L}_m, T) \) gravity, where the black hole is surrounded by an anisotropic fluid acting as the matter source. We consider both linear and nonlinear forms of the function \( f(R, \mathcal{L}_m, T) \), enabling a detailed exploration of how anisotropic pressures and different \( f(R, \mathcal{L}_m, T) \) influence the spacetime structure. Furthermore, we derive the conditions on the coupling parameters \( (\beta_1, \beta_2, \beta_3) \) under which the energy conditions are satisfied. Utilising these constraints, we then investigate the thermodynamic behaviour of the resulting black hole solutions in the presence of various matter fields, namely, dust, radiation, and a quintessence field, each characterised by a distinct equation-of-state parameter. An important outcome of this study is that the results obtained deviate from those predicted by standard General Relativity. It is also observed that these deviations depend explicitly on the interaction between the matter Lagrangian \( \mathcal{L}_m \) and the trace \( T \) of the energy-momentum tensor.
\par
\vspace{0.1cm}
\textbf{Keyword:} Black Hole; Modified gravity; Thermodynamic.
\end{abstract}

\maketitle
\section{Introduction}\label{sec1}
The discovery of the universe's accelerated expansion in the late 1990s \cite{I1,I2,I3,I4,I5} prompted extensive research efforts aimed at understanding this phenomenon, as it could not be satisfactorily explained within the framework of General Relativity. In response, two primary approaches have been proposed to account for this late-time cosmic acceleration:  
(i) introducing additional matter and energy components—commonly referred to as the dark constituents of the universe—into the framework of general relativity; and  
(ii) modifying or extending the Einstein-Hilbert action by allowing the gravitational Lagrangian to be an arbitrary function of geometric and/or matter-related quantities. In this paper, we focus on the modification of the Einstein-Hilbert action as an approach to address the limitations of General Relativity.However, these modified gravity theories are challenging to analyze due to the presence of higher-order derivative terms in their equations of motion.
\par
The motivation for developing modified theories of gravity stems from several critical limitations in our current understanding of gravitational phenomena \cite{I6,I7,I8}. (i) Unification of Physics: General Relativity (GR) provides an excellent description of gravity at macroscopic scales but fails to reconcile with quantum mechanics, which governs physics at microscopic scales. Bridging this gap and formulating a unified framework—commonly referred to as quantum gravity—is a major objective in theoretical physics. Modifying GR is often considered a plausible route toward achieving this unification.
   (ii)Dark Matter and Dark Energy: Observational evidence supports the existence of dark matter and dark energy \cite{I9,I10,I11,I12}, entities that do not fit comfortably within the framework of GR. Dark matter \cite{I13,I14,I15}, for example, influences the gravitational dynamics of galaxies, while dark energy is believed to be driving the observed accelerated expansion of the universe. Modified gravity theories offer a promising avenue for explaining these phenomena without invoking unknown forms of matter or energy.
  (iii)Inconsistencies in Observations: There are instances where astrophysical and cosmological observations do not fully align with the predictions of General Relativity. These discrepancies may suggest that GR is incomplete or that additional modifications are necessary to provide a more accurate description of gravitational interactions on large scales. Modified gravity theories are therefore explored as potential frameworks for resolving such inconsistencies and for offering improved explanations of certain observational data. While GR remains remarkably successful in describing a broad spectrum of gravitational phenomena, extending or modifying the theory could lead to a deeper and more unified understanding of the universe.
\par
The development of modified theories of gravity typically begins by introducing correction terms to the Einstein–Hilbert action. These additional terms aim to capture deviations from General Relativity and address various cosmological and astrophysical challenges. Notable examples include \( f(R) \) gravity \cite{I16,I17,I18}, \( f(G) \) gravity \cite{I19,I20},$f(P)$ gravity \cite{I45}, \( f(\mathcal T) \) gravity \cite{I21,I22,I23,I24}, and the more generalized \( f(\mathcal T, T_G) \) gravity \cite{I25,I26}. Other significant frameworks include Weyl gravity\cite{I27,I28} and Lovelock gravity \cite{I29,I30}, among others. Each of these theories extends the geometrical structure of gravity in a unique way, offering new insights into the behaviour of the spacetime.
\par
In this study, we investigate the framework of \( f(R, \mathcal{L}_m, T) \) gravity, a generalised theory that extends standard General Relativity by allowing a non-minimal interaction between matter fields and spacetime curvature. This approach, often referred to as the ``dark coupling scenario,'' has been proposed as a viable alternative to explain the universe’s late-time acceleration without invoking dark energy explicitly~\cite{I31}.
 Broadly, three major frameworks have been proposed to explain this phenomenon:
(i) The dark components model extends Einstein’s field equations by incorporating supplementary contributions into the total energy-momentum tensor, effectively allowing for a distinct treatment of dark matter and dark energy components; 
(ii) the dark gravity framework, which interprets cosmic acceleration through purely geometric modifications of gravity; and  
(iii) the dark coupling approach, which generalises the gravitational Lagrangian by intricately coupling matter and geometry.
The central idea of the dark coupling framework is to replace the conventional Einstein-Hilbert Lagrangian—where the curvature and matter terms appear as additive, independent contributions—with a more general functional form that incorporates algebraic coupling between these components. In \( f(R, \mathcal{L}_m, T) \) gravity, the gravitational Lagrangian is an arbitrary function of the Ricci scalar \( R \), the matter Lagrangian \( \mathcal{L}_m \), and the trace \( T \) of the energy-momentum tensor. This naturally leads to a non-minimal interaction between matter and spacetime geometry, enabling a rich structure of coupling mechanisms, such as those involving the Ricci scalar and \( T \), which are absent in General Relativity. As special cases, the \( f(R, \mathcal{L}_m, T) \) gravity framework reduces to several well-known modified gravity models: it recovers \( f(R) \) gravity when the dependence on \( \mathcal{L}_m \) and \( T \) is removed; it yields \( f(R, \mathcal{L}_m) \) gravity when the trace \( T \) is excluded; and it corresponds to \( f(R, T) \) gravity in the absence of explicit \( \mathcal{L}_m \) dependence.
\par
In this paper, we derive the black hole metric function in the presence of an anisotropic fluid within the framework of modified \( f(R, \mathcal{L}_m, T) \) gravity, considering both linear and nonlinear forms of the function \( f(R, \mathcal{L}_m, T) \). We further investigate the associated thermodynamic properties of the resulting black hole solutions. Some examples of black holes in other modified gravity theories, such as \( f(R) \) gravity\cite{I32,I33,I34}, \( f(G) \) gravity\cite{I35,I36}, and \( f(T) \) gravity \cite{I37,I38,I39}, and so on.
\par
The structure of the paper is as follows: In Section~\ref{sec2}, we formulate the field equations for the modified \( f(R, \mathcal{L}_m, T) \) gravity. In Section~3, we obtain the black hole metric function \( \psi(r) \) and analyze its behavior along with the corresponding Hawking temperature for fixed values of the equation of state parameter \( w = 0, \frac{1}{3}, -\frac{2}{3} \), representing dust, radiation, and quintessence fields, respectively. In Section~4, we extend our analysis to a nonlinear form of \( f(R, \mathcal{L}_m, T) \), derive the associated metric function, and investigate its characteristics and thermodynamic behaviour for the same choices of \( w \).

 \section{FIELD EQUATION OF $f(R,\mathcal L_{m},T)$ GRAVITY}\label{sec2}
 To derive the field equations within the framework of a modified theory of gravity, we begin by considering the gravitational action, which is expressed as follows:
\begin{equation} \label{e1}
      I=\int \sqrt{-g}\left(\frac{1}{2k}f(R,\mathcal L_{m},T)+\mathcal L_{m}\right)\ d^4x  
    \end{equation}
   Here, \( f(R, \mathcal{L}_m, T) \) denotes an arbitrary function of the Ricci scalar \( R \), the matter Lagrangian density \( \mathcal{L}_m \),$k$ is  Einstein gravitational constant  and the trace \( T \) of the stress-energy tensor,\( T_{\mu\nu} \). The stress-energy tensor is defined by the variational relation as:
    \begin{equation}\label{e2}
T_{\mu\nu} = -2 \sqrt{-g} \, \frac{\partial (\sqrt{-g} \, \mathcal L_m)}{\partial g^{\mu\nu}} = -2 \frac{\partial\mathcal L_m}{\partial g^{\mu\nu}} + g_{\mu\nu}\mathcal L_m
 \end{equation}
 In this work, it is assumed that the matter Lagrangian \( \mathcal{L}_m \) is a function solely of the metric tensor \( g_{\mu\nu} \), and does not involve derivatives of the metric.
Now,Varying the action~(\ref{e1}) with respect to the metric \( g^{\mu\nu} \), we obtain the following expression:

    \begin{equation}\label{e3}
\begin{split}
      \delta I =& \frac{1}{2\kappa} \int \Big[ f_R \delta R + f_ {\mathcal L_{m}} \delta \mathcal L_{m} + f_T \delta T + f \delta \left( \sqrt{-g} \right) \\
      &+ \frac{2\kappa}{\sqrt{-g}} \delta \left( \mathcal{L}_m \sqrt{-g} \right) \Big]\sqrt{-g}d^4x
\end{split}
\end{equation}
Here, \( f_R \), \( f_{\mathcal{L}_m} \), and \( f_T \) denote the partial derivatives of the function \( f(R, \mathcal{L}_m, T) \) with respect to \( R \), \( \mathcal{L}_m \), and \( T \), respectively.
The variations of the first three terms in Eq.~(\ref{e3}) are computed as follows:
 \begin{equation}\label{e4}
  f_{R} \delta R = (R_{\mu\nu} + g_{\mu\nu} \nabla^2 - \nabla_\mu \nabla_\nu )f_{R}\delta g^{\mu\nu} 
\end{equation}
\begin{equation}\label{e5}
  f_{T} \delta T = (T_{\mu\nu}  + \Theta_{\mu\nu} )f_{T}\delta g^{\mu\nu} 
 \end{equation}
 \begin{equation}\label{e6}
     f_{ \mathcal L_{m}}\delta \mathcal L_{m}=\frac{(g_{\mu \nu}\mathcal L_{m}-T_{\mu \nu}) f_{ \mathcal L_{m}} \delta g^{\mu\nu}}{2}
 \end{equation}
 Where $\Theta_{\mu\nu}$ defined  as:
 \begin{equation}\label{e7}
\Theta_{\mu\nu} =\frac{ \delta T_{\alpha\beta}}{\delta g^{\mu\nu} } g^{\alpha\beta} = -2 T_{\mu\nu} + g_{\mu\nu} \mathcal L_m - \eta_{\mu\nu}
\end{equation}
Here, we introduce a tensor \( \eta_{\mu\nu} \), defined as:
\begin{equation}\label{e8}
\eta_{\mu\nu} = 2 g^{\alpha\beta} \frac{\partial^2 L_m}{\partial g^{\mu\nu} \partial g^{\alpha\beta}}
\end{equation}
By imposing the condition \( \delta I = 0 \) in Eq.~(\ref{e3}) and utilizing Eqs.~(\ref{e4}) through (\ref{e6}), we obtain the field equations for \( f(R, \mathcal{L}_m, T) \) gravity as:
 \begin{equation} \label{e9}
 \begin{split}
f_R R_{\mu\nu} - \frac{1}{2} \left[ f - (f_L + 2f_T)L_m \right] g_{\mu\nu} + \left( g_{\mu\nu} \nabla^2 - \nabla_{\mu} \nabla_{\nu} \right) f_R 
\\= \left[ \kappa + \frac{1}{2}(f_L + 2f_T) \right] T_{\mu\nu} + f_T \eta_{\mu\nu}
\end{split}
 \end{equation}
Here, \( \nabla^2 \) denotes the d'Alembert operator, defined as \( \nabla^2 = \nabla^\mu \nabla_\mu \), where \( \nabla_\mu \) represents the covariant derivative associated with the metric. By considering \( f(R, \mathcal{L}_m, T) = f(R) \), where the function is independent of both \( \mathcal{L}_m \) and \( T \), we recover the field equations of \( f(R) \) gravity, as originally proposed in Ref. \cite{I40}. Furthermore, if the function takes the form \( f(R, \mathcal{L}_m, T) = f(R, T) \) or \( f(R, \mathcal{L}_m, T) = f(R, \mathcal{L}_m) \), we obtain the corresponding field equations of \( f(R, T) \) and \( f(R, \mathcal{L}_m) \) gravity, respectively, as discussed in Refs.~\cite{I41,I42}.Taking the covariant derivative of Eq.(~\ref{e9}), we obtain the following expression for the divergence of the stress-energy tensor \( T_{\mu\nu} \):
 \begin{equation}\label{e10}
 \begin{split}
  \nabla ^{\mu} T_{\mu \nu}= &\frac{1}{k+f_{t}}\Big[ g_{\mu,\nu} \nabla^{\mu} (f_{t} \mathcal L_m)-T_{\mu,\nu} \nabla ^{\mu }(f_{t})- \nabla ^{\mu }(f_{T}\eta_{\mu,\nu})\\
&  -\frac{1}{2}(f_{T}\nabla^{\mu}T+f_{\mathcal L_{m}}\nabla^{\mu}\mathcal L_m)\Big]    
 \end{split}
\end{equation}
Here, we define \( f_t \) as follows:
\begin{equation}\label{e11}
    f_t=f_T+\frac{1}{2}f_{\mathcal L_{m}}
\end{equation}
Next, we can evaluate the tensor \( \Theta_{\mu\nu} \), as defined in Eq.~(\ref{e7}), once the explicit form of the matter Lagrangian \( \mathcal{L}_m \) is specified.In the present analysis, we assume that the stress-energy tensor of the matter is given by:
\begin{equation}\label{e12}
T^{\mu\nu} = (\rho + p_t) u^\mu u^\nu + p_t g_{\mu\nu}+(p_r-p_t)v^\mu v^\nu
\end{equation}
Here, \( p_r \) and \( p_t \) denote the radial and tangential pressures of the anisotropic fluid, respectively. In this study, we adopt the relations \( p_r = -\rho \) and \( p_t = \frac{1}{2}\rho(3\omega + 1) \), where \( \rho \) is the energy density and \( \omega \) is a state parameter.
The four-velocity \( u^\mu \) satisfies the condition \( u^\mu u_\mu = 1 \), and \( u^\mu \nabla _\nu u_\mu = 0 \).We now consider the matter Lagrangian density associated with an anisotropic fluid to be \( \mathcal{L}_m = -\frac{1}{3}(p_r + 2p_t) \)\cite{I43}. Then, using Eq.~(\ref{e7}), we obtain the variation of the stress-energy tensor for the anisotropic fluid in the following form:
\begin{equation}\label{e13}
    \Theta_{\mu\nu} = -2 T_{\mu\nu} - \frac{1}{3}(p_r + 2p_t) g_{\mu\nu}.
\end{equation}
In the following section, we consider specific classes of \( f(R, \mathcal{L}_m, T) \) modified gravity models by explicitly specifying the functional form of \( f \). For each chosen form, we solve the corresponding modified field equations using the anisotropic matter distribution discussed above. Subsequently, we obtain the black hole metric solutions and investigate their associated thermodynamic properties.
\section{Black Hole solution for Liner function of $f(R,\mathcal L_{m},T)$ }\label{Sec3}
As a first case of an \( f(R, \mathcal{L}_m, T) \) modified gravity model, we consider a linear functional form given by$f(R, \mathcal{L}_m, T) = \beta_1 R + \beta_2 \mathcal{L}_m + \beta_3 T$,
where \( \beta_1 \), \( \beta_2 \), and \( \beta_3 \) coupling parameters.
The gravitational field equations corresponding to the linear form of \( f(R, \mathcal{L}_m, T) \), as given in Eq.~(\ref{e9}), follow directly and take the form:
\begin{equation}\label{e14}
  \begin{split}
G_{\mu \nu}&=\frac{(k+\beta)}{\beta_{1}}T_{\mu \nu}+\frac{1}{2\beta_{1}}(\beta_{2}\mathcal L_{m}+\beta_{3}T)g_{\mu \nu} -\frac{\beta}{\beta_{1}} \mathcal L_{m} g_{\mu \nu}\\
&=\tilde{T_{\mu\nu}}
  \end{split}
\end{equation}
where we defined  $\beta=\frac{1}{2}(\beta_{2}+2\beta_{3})$.
We now assume a static and spherically symmetric spacetime, described by the following line element:
\begin{equation}\label{e15}
ds^2 = -\psi(r) \, dt^2 + \frac{dr^2}{\psi(r)} + r^2 \left( d\theta^2 + \sin^2\theta \, d\phi^2 \right)
\end{equation}
By substituting the static, spherically symmetric metric into the modified field equations [Eq.~(\ref{e9})], we obtain two independent components: \( G^t_{\ t} = G^r_{\ r} \) and \( G^\theta_{\ \theta} = G^\phi_{\ \phi} \). These yield two independent field equations, which lead to the following differential equations:

\begin{equation}\label{e16}
\frac{1}{2} \Big( 2k + \beta_2 + (3 - 5w) \beta_3 \Big) \rho(r) + \frac{\beta_1 \Big( -1 + \psi(r) + r \psi'(r) \Big)}{r^2} = 0
\end{equation}
\begin{equation}\label{e17}
-\frac{1}{4} (1 + 3w) (2k + \beta_2) \rho(r) - 4w \beta_3 \rho(r) + \frac{\beta_1 \psi'(r)}{r} + \frac{1}{2} \beta_1 \psi''(r) = 0
\end{equation}
Thus, we are left with two unknown functions, \( \psi(r) \) and \( \rho(r) \), which can be determined analytically by solving the two independent field equations. Eliminating \( \rho(r) \) from Eqs.~(\ref{e16}) and (\ref{e17}), and subsequently solving the resulting differential equation, we obtain the metric function \( \psi(r) \).
\begin{widetext}
\begin{equation}\label{e18}
 \psi(r)=1 + \frac{c_2}{r} - \frac{\left(2k + \beta_2 + (3 - 5w)\beta_3\right)c_1 r^{-1 - \frac{3\left(2k w - \beta_3 + w(\beta_2 + 7\beta_3)\right)}{2k + \beta_2 + (3 - 5w)\beta_3}}}{3\left(2k w - \beta_3 + w(\beta_2 + 7\beta_3)\right)}
\end{equation}
\end{widetext}
Now, by substituting the solution for \( \psi(r) \) into the field equation, we can obtain the corresponding energy density \( \rho(r) \).
\begin{equation}\label{e19}
    \rho(r) = -\frac{2 \beta_1 c_1 \, r^{-\frac{3 (1 + w) (2 k + \beta_2 + 2 \beta_3)}{2 k + \beta_2 + (3 - 5w) \beta_3}}}{2 k + \beta_2 + (3 - 5w) \beta_3}
\end{equation}
Next, we want to determine the values of the integration constants \( c_1 \) and \( c_2 \). We consider a limiting case of the theory. Specifically, we take the parameters \( \beta_2 \to 0 \), \( \beta_3 \to 0 \), and assume an equation of state parameter \( w = -1 \), corresponding to a  cosmological constant field. Additionally, we set \( \beta_1 = 1 \), such that the gravitational sector effectively reduces to general relativity in this limit.

Under these assumptions, the metric function \( \psi(r) \) simplifies to the form:
\[
\psi(r) = 1 + \frac{c_2}{r} + \frac{c_1 r^2}{3}.
\]

This form is immediately recognizable as the general solution for a static, spherically symmetric spacetime in the presence of a cosmological constant, namely the Schwarzschild–de Sitter metric. By comparing this result with the standard form:
\[
\psi(r) = 1 - \frac{2M}{r} - \frac{\Lambda r^2}{3},
\]
We identify the constants as follows:
\[
c_1 = -\Lambda, \quad c_2 = -2M,
\]
where \( M \) represents the total mass of the black hole and \( \Lambda \) denotes the cosmological constant. Finally, the resulting  metric becomes:
\begin{widetext}
\begin{equation}\label{20}
 \psi(r)=1 - \frac{2M}{r} + \frac{\left(2k + \beta_2 + (3 - 5w)\beta_3\right)\Lambda \, r^{-1 - \frac{3\left(2k w - \beta_3 + w(\beta_2 + 7\beta_3)\right)}{2k + \beta_2 + (3 - 5w)\beta_3}}}{3\left(2k w - \beta_3 + w(\beta_2 + 7\beta_3)\right)}
\end{equation}
\end{widetext}
\subsection{Energy Condition}
\begin{figure}[H]
    \centering
    \includegraphics[width=1\linewidth]{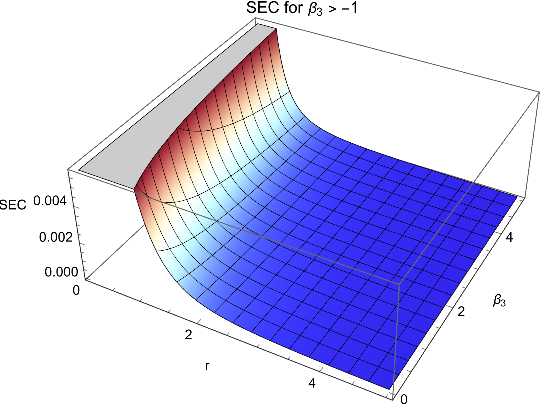}
    \caption{For the dust field (\( w = 0 \)), the strong energy condition (SEC) is satisfied when the parameter \( \beta_3 > -1 \).}
    \label{f1}
    \end{figure}

\begin{figure}[H]
    \centering
    \includegraphics[width=1\linewidth]{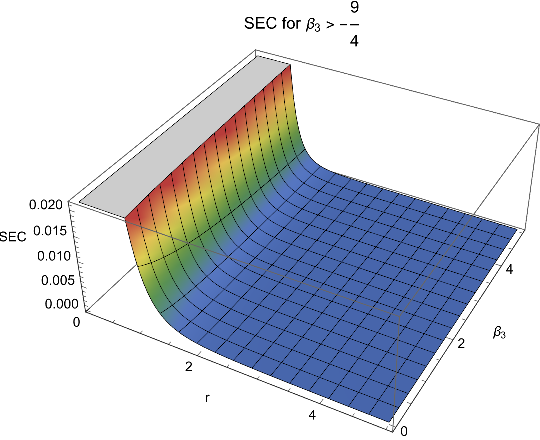}
    \caption{For the radiation field (\( w = 1/3 \)), the strong energy condition (SEC) is satisfied when the parameter \( \beta_3 < -\frac{9}{4} \).   }
    \label{f2}
    \end{figure}

\begin{figure}[H]
    \centering
    \includegraphics[width=1\linewidth]{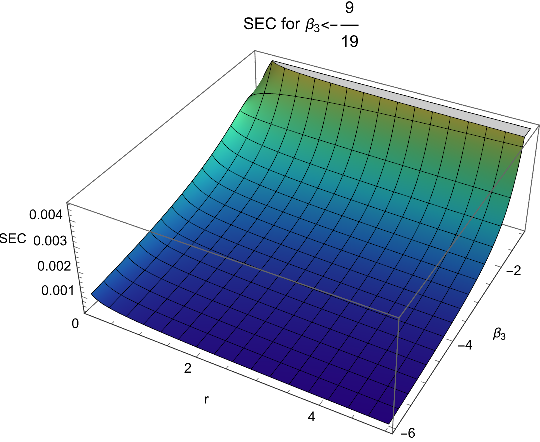}
    \caption{For the quintessence field (\( w = -2/3 \)), the strong energy condition (SEC) is satisfied when the parameter \( \beta_3 < -\frac{9}{19} \).}
    \label{f3}
    \end{figure}

Energy conditions impose important constraints on the components of the energy-momentum tensor to ensure that the matter content is physically reasonable. For anisotropic fluids, which exhibit different pressures along radial and tangential directions, these conditions play a critical role in characterizing the matter distribution. In particular, the Strong Energy Condition (SEC) for anisotropic matter is given by
\begin{equation}\label{e21}
    \text{SEC:} \quad \rho + p_n \geq 0, \quad \rho + \sum_n p_n \geq 0
\end{equation}
where $n=1,2,3...$.Therefore, the SEC can written as
\begin{equation}\label{e22}
\begin{split}
 \rho + p_r +  2p_t=\frac{2(1 + 3w) \beta_1 \Lambda r^{-\frac{3(1 + w)(2k + \beta_2 + 2\beta_3)}{2k + \beta_2 + (3 - 5w)\beta_3}}}{2k + \beta_2 + (3 - 5w)\beta_3}
 \end{split}
\end{equation}
From Eqs.~(\ref{e21}) and (\ref{e22}), the criteria ensuring the validity of the strong energy condition (SEC) for the anisotropic fluid configuration can be expressed as:
\begin{equation}\label{e23}
\frac{2(1 + 3w) \beta_1}{2k + \beta_2 + (3 - 5w)\beta_3} \geq 0
\end{equation}
Therefore, the admissible set of solutions that satisfy the above inequality can be expressed in terms of the model parameters \( \beta_1 \) , \( \beta_2 \) and state parameter $w$ , with the constant \( k = 1 \), as follows:
\begin{widetext}
\begin{equation}\label{e24}
S = 
\begin{cases} 
w < -\frac{1}{3}, & \left( \beta_3 < \frac{2 + \beta_2}{-3 + 5w} \land \beta_1 \geq 0 \right) \lor \left( \beta_3 > \frac{2 + \beta_2}{-3 + 5w} \land \beta_1 \leq 0 \right) \\
w = -\frac{1}{3}, & \left( \beta_3 < -\frac{3}{14}(2 + \beta_2) \right) \lor \left( \beta_3 > -\frac{3}{14}(2 + \beta_2) \right) \\
-\frac{1}{3} < w < \frac{3}{5}, & \left( \beta_3 < \frac{2 + \beta_2}{-3 + 5w} \land \beta_1 \leq 0 \right) \lor \left( \beta_3 > \frac{2 + \beta_2}{-3 + 5w} \land \beta_1 \geq 0 \right) \\
w = \frac{3}{5}, & \left( \beta_2 < -2 \land \beta_1 \leq 0 \right) \lor \left( \beta_2 > -2 \land \beta_1 \geq 0 \right) \\
w > \frac{3}{5}, & \left( \beta_3 < \frac{2 + \beta_2}{-3 + 5w} \land \beta_1 \geq 0 \right) \lor \left( \beta_3 > \frac{2 + \beta_2}{-3 + 5w} \land \beta_1 \leq 0 \right)
\end{cases}
\end{equation}
\end{widetext}
Where \( \lor \) and \( \land \) represent the logical OR and AND operators, respectively. Equation~(\ref{e24}) provides the complete set of solutions to the inequality presented in Equation~(\ref{e23}).In this work, we perform a detailed analysis of the strong energy condition (SEC) for a fixed value of the equation-of-state parameter \( w \). Specifically, we fix the model parameter \( \beta_2 = 1 \) and investigate how the satisfaction of the SEC depends on the variation of the parameter \( \beta_3 \). This allows us to identify the range of \( \beta_3 \) values for which the SEC holds, thereby constraining the viable parameter space of the model under consideration. The conditions \( \beta_3 > -1 \), \( \beta_3 > -\frac{9}{4} \), and \( \beta_3 < -\frac{9}{19} \) represent the values of \( \beta_3 \) for which the strong energy condition (SEC) is satisfied in the case of a black hole surrounded by different matter fields: dust (\( w = 0 \)), radiation (\( w = \frac{1}{3} \)), and quintessence (\( w = -\frac{2}{3} \)), respectively. The corresponding behaviour of the SEC can also be visualised in Figures~\ref{f1}, \ref{f2}, and \ref{f3}. In the next subsection, we determine the black hole mass and Hawking temperature, and analyze the thermodynamic behavior of the black hole surrounded by different matter fields: dust (\( w = 0 \)), radiation (\( w = 1/3 \)), and a quintessence field (\( w = -2/3 \)), using the SEC conditions established in this subsection.
\subsection{PARTICULAR CASE}
In this subsection, we present a detailed analysis of the black hole metric structure, compute the black hole mass, and examine the thermodynamic behavior of black holes surrounded by different matter fields: dust (\( w = 0 \)), radiation (\( w = 1/3 \)), and a quintessence field (\( w = -2/3 \)).The black hole metric in the context of the modified \( f(R, \mathcal{L}_m, T) \) gravity theory is given as follows:
\begin{equation}\label{e25}
\begin{split}
ds^2 &=\\
&\small{-\Big(1 - \frac{2M}{r} + \frac{\left(2k + \beta_2 + (3 - 5w)\beta_3\right)\Lambda \, r^{-1 - \frac{3\left(2k w - \beta_3 + w(\beta_2 + 7\beta_3)\right)}{2k + \beta_2 + (3 - 5w)\beta_3}}}{3\left(2k w - \beta_3 + w(\beta_2 + 7\beta_3)\right)}\Big) \, dt^2 }\\
&+ \frac{dr^2}{\Big(1 - \frac{2M}{r} + \frac{\left(2k + \beta_2 + (3 - 5w)\beta_3\right)\Lambda \, r^{-1 - \frac{3\left(2k w - \beta_3 + w(\beta_2 + 7\beta_3)\right)}{2k + \beta_2 + (3 - 5w)\beta_3}}}{3\left(2k w - \beta_3 + w(\beta_2 + 7\beta_3)\right)}\Big)} \\
&+ r^2 \left( d\theta^2 + \sin^2\theta \, d\phi^2 \right)
\end{split}
\end{equation}
where \( M \) represents the total mass of the black hole and \( \Lambda \) denotes the cosmological constant.Let \( r_h \) denote the radial coordinate at which the metric function \( \psi(r) \) vanishes, i.e., \( \psi(r_h) = 0 \); this defines the location of the event horizon. Utilizing this condition, the black hole mass can be expressed as a function of the horizon radius \( r_h \), as follows:
\begin{equation}\label{e26}
    M(r_h)=\frac{r_h}{2} + 
\frac{r_h^{\left( \frac{3 \beta_3 - 3 w (2k + \beta_2 + 7\beta_3)}{2k + \beta_2 + 3\beta_3 - 5w \beta_3} \right)} (2k + \beta_2 + 3\beta_3 - 5w \beta_3)  \Lambda}{-6 \beta_3 + 6w (2k + \beta_2 + 7\beta_3)}
\end{equation}
This relation provides a general expression connecting the black hole mass with the state parameter \( w \), as well as the coupling constants and \( \Lambda \). We observe that, in \( f(R, \mathcal{L}_m, T) \) gravity, the mass of a black hole surrounded by a fluid depends not only on the horizon radius but also on the state parameter \( w \), the coupling constants, and the cosmological constant. On the other hand, the Hawking temperature~\cite{I44} associated with our solution can be expressed in terms of the horizon radius as follows:
\begin{equation}\label{e27}
\begin{split}
\textbf{ T}=\frac{  r_h^{- \left( \frac{3 (1 + w) (2k + \beta_2 + 2\beta_3)}{2k + \beta_2 + (3 - 5w)\beta_3} \right)}\left( 
r_h^{\left( \frac{(2 + 3w)(2k + \beta_2) + (3 + 11w)\beta_3}{2k + \beta_2 + 3\beta_3 - 5w\beta_3} \right)} 
- r_h \Lambda  \right)}{4\pi}
\end{split}
\end{equation}
From this expression, we observe that the Hawking temperature of a black hole surrounded by an anisotropic fluid in \( f(R, \mathcal{L}_m, T) \) gravity exhibits an additional structure arising from its dependence on the state parameter \( w \), the coupling constants, and the cosmological constant \( \Lambda \).In the following sub-subsection, we analyse the implications of this result by considering specific choices of the state parameter \( w \).
\subsubsection{Black hole surrounded by a dust field}
To investigate the thermodynamic properties of the black hole in a more concrete setting, we consider the case of a dust-like surrounding field characterised by \( w = 0 \). Additionally, we fix constant \( k = 1 \), and set the coupling constant \( \beta_2 = 1 \) for simplicity. Under these assumptions, the metric function \( \psi(r) \)(\ref{e28}) and the associated Hawking temperature (\ref{e29}) acquire simplified forms, which facilitate further analysis of the gravitational and thermodynamic behavior in the context of \( f(R, \mathcal{L}_m, T) \) gravity.
\begin{equation}\label{e28}
\begin{split}
 ds^2 &= -\Big(1 - \frac{2M}{r} - \frac{
    r^{- \left( \frac{1}{1 + \beta_3} \right)} (1 + \beta_3) \Lambda}{\beta_3}
\Big) dt^2 \\
&+ \frac{dr^2}{\Big(1 - \frac{2M}{r} - \frac{
    r^{- \left( \frac{1}{1 + \beta_3} \right)} (1 + \beta_3) \Lambda}{\beta_3}\Big) } \\
&+ r^2 \left( d\theta^2 + \sin^2\theta d\phi^2 \right)   
\end{split}
\end{equation}
\begin{equation}\label{e29}
\textbf{T}= \frac{ 1 - r_h^{- \left( \frac{1}{1 + \beta_3} \right)} \Lambda}{ 4\pi r_h}
\end{equation}
Figure~\ref{f4} illustrates the behavior of the metric function \( \psi(r) \) for a black hole surrounded by a dust field in the framework of \( f(R, \mathcal{L}_m, T) \) gravity, with the cosmological constant fixed at \( \Lambda = 0.001 \). The profiles are plotted for selected values of the parameter \( \beta_3 \), which are chosen to ensure that the strong energy condition (SEC) is satisfied. These admissible values of \( \beta_3 \) are determined based on the constraints outlined in Subsection~A.
The figure (\ref{f4}) illustrates the behavior of the metric function \( \psi(r) \) for various values of the parameter \( \beta_3 \) in \( f(R, \mathcal{L}_m, T) \) gravity, with the cosmological constant fixed at \( \Lambda \). It is evident that the variation in \( \beta_3 \) produces only minor deviations in the profile of \( \psi(r) \). Notably, all curves asymptotically approach the Schwarzschild solution, and in the limit of large \( r \), the metric function tends to unity, i.e., \( \psi(r) \rightarrow 1 \).
\par
Equation (\ref{e29}) represents the Hawking temperature associated with a black hole surrounded by a dust matter distribution in the context of \( f(R, \mathcal{L}_m, T) \) gravity. The expression encapsulates the dependence of the black hole temperature on both the horizon radius \( r_h \) and the matter-curvature coupling parameter \( \beta_3 \), resulting in a family of temperature profiles. Figure~\eqref{f5} illustrates the variation of the Hawking temperature {\( \textbf{T} \) as a function of the event horizon radius for several representative values of \( \beta_3 \).
From the plots, it is evident that the Hawking temperature remains strictly positive for all configurations considered. Moreover, a clear trend emerges wherein the temperature increases as the horizon radius \( r_h \) decreases. The dependence on \( \beta_3 \) reflects the influence of the non-minimal coupling between matter and geometry, subtly modifying the thermodynamic properties of the black hole without introducing unphysical behaviour such as negative temperatures.
\begin{figure}[H]
    \centering
    \includegraphics[width=1\linewidth]{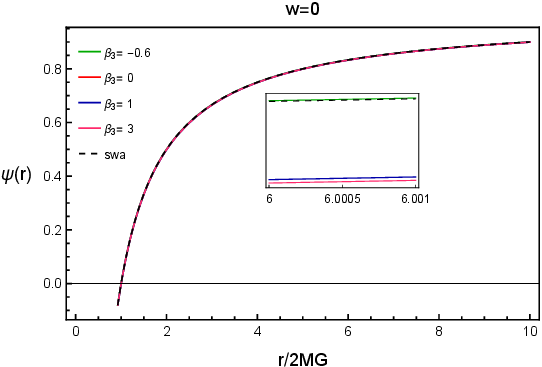}
    \caption{The behaviour of \( \psi(r) \) is analysed for different values of \( \beta_3 \), assuming a dust matter distribution and a fixed cosmological constant \( \Lambda = 0.001 \).}
    \label{f4}
    \end{figure}
    \begin{figure}[H]
    \centering
    \includegraphics[width=1\linewidth]{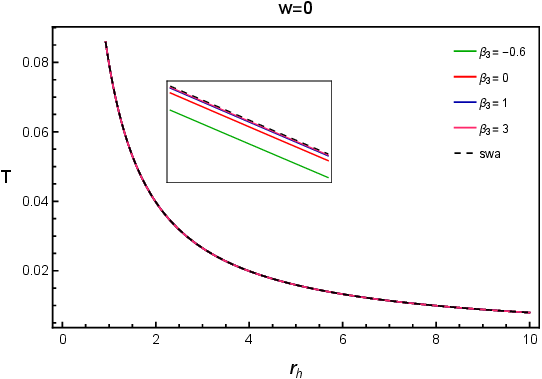}
    \caption{Equation (30) is graphically represented as a function of the horizon radius \( r_h \) and the coupling parameter \( \beta_3 \), for a fixed cosmological constant \( \Lambda = 0.001 \) and an equation of state parameter \( w = 0 \) corresponding to a dust fluid. }
    \label{f5}
    \end{figure} 
    
\subsubsection{Black hole surrounded by a radiation field}
In the second case, we examine the properties of a black hole surrounded by a radiation field, characterized by an equation of state parameter \( w = 1/3 \), with \( \beta_2 = 1 \) and \( k = 1 \), within the framework of the modified \( f(R, \mathcal{L}_m, T) \) gravity theory. In this scenario, the black hole metric and the corresponding Hawking temperature are obtained in closed form and are given by the following expressions:
\begin{equation}\label{e30}
\begin{split}
 ds^2 &= -\Big(1 - \frac{2M}{r} + \frac{
    r^{-4 + \frac{18}{9 + 4\beta_3}} (9 + 4\beta_3) \Lambda}{3(3 + 4\beta_3)}
\Big) dt^2 \\
&+ \frac{dr^2}{\Big(1 - \frac{2M}{r} + \frac{
    r^{-4 + \frac{18}{9 + 4\beta_3}} (9 + 4\beta_3) \Lambda}{3(3 + 4\beta_3)}
\Big) } \\
&+ r^2 \left( d\theta^2 + \sin^2\theta d\phi^2 \right)   
\end{split}
\end{equation}
\begin{equation}\label{e31}
  \textbf{ T}=\frac{
    r_h^4 - r_h^{\frac{18}{9 + 4\beta_3}} \Lambda}{4\pi r_h^5}
\end{equation}
Figure~\eqref{f6} depicts the behavior of the metric function \( \psi(r) \) for a black hole surrounded by a radiation field (\( w = 1/3 \)) in the context of \( f(R, \mathcal{L}_m, T) \) gravity, with a fixed cosmological constant \( \Lambda \). The curves correspond to different values of the parameter \( \beta_3 \). It is observed that all metric profiles converge to the Schwarzschild solution at large radial distances, where \( \psi(r) \to 1 \). Near the black hole, however, small deviations from general relativity arise due to the effects of the modified gravity terms.
\par
Equation~\eqref{e31} describes the Hawking temperature of a black hole surrounded by a radiation field within the framework of \( f(R, \mathcal{L}_m, T) \) gravity. As illustrated in Fig.\eqref{f6}, the temperature remains positive for all values of the horizon radius \( r_h \), exhibiting an inverse relationship where the temperature increases as \( r_h \) decreases.
\begin{figure}[H]
    \centering
    \includegraphics[width=1\linewidth]{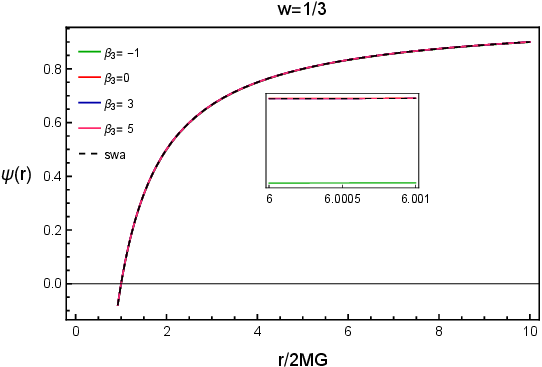}
    \caption{The behaviour of \( \psi(r) \) is analysed for different values of \( \beta_3 \), assuming $w=1/3$ and a fixed cosmological constant \( \Lambda = 0.001 \).}
    \label{f6}
    \end{figure}
    \begin{figure}[H]
    \centering
    \includegraphics[width=1\linewidth]{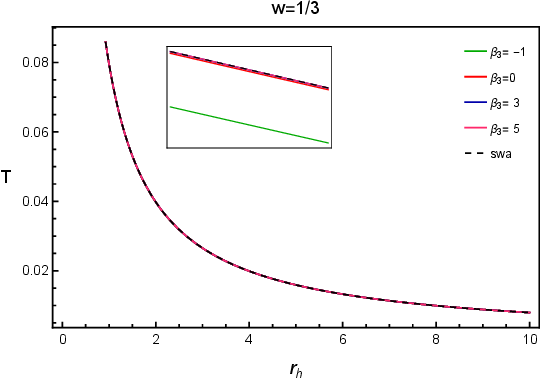}
    \caption{Equation \eqref{e31} is graphically represented as a function of the horizon radius \( r_h \) and the coupling parameter \( \beta_3 \), for a fixed cosmological constant \( \Lambda = 0.001 \) and an equation of state parameter \( w = 1/3 \) corresponding to a radiation fluid. }
    \label{f7}
    \end{figure} 
\subsubsection{Black hole surrounded by a quintessence  Field } 
As a third case, we investigate the properties of a black hole surrounded by a quintessence field within the framework of modified \( f(R, \mathcal{L}_m, T) \) gravity. The quintessence field is characterized by an equation of state parameter \( w = -\frac{2}{3} \), and we fix the parameters \( \beta_2 = 1 \) and \( k = 1 \). Under these conditions, we derive closed-form expressions for both the black hole metric \eqref{e32} and the corresponding Hawking temperature\eqref{e33}. These solutions provide insight into how the presence of quintessence modifies the spacetime geometry and thermodynamic properties of the black hole.
\begin{equation}\label{e32}
\begin{split}
 ds^2 &= -\Big(1 - \frac{2M}{r} - \frac{
    r^{\frac{9 + 32\beta_3}{9 + 19\beta_3}} (9 + 19\beta_3) \Lambda
}{ 18 + 51\beta_3}\Big) dt^2 \\
&+ \frac{dr^2}{\Big(1 - \frac{2M}{r} - \frac{
    r^{\frac{9 + 32\beta_3}{9 + 19\beta_3}} (9 + 19\beta_3) \Lambda}{  18 + 51\beta_3}
\Big) } \\
&+ r^2 \left( d\theta^2 + \sin^2\theta d\phi^2 \right)   
\end{split}
\end{equation}
\begin{equation}\label{e33}
  \textbf{ T}=\frac{ 1 - r_h^{\frac{9 + 32\beta_3}{9 + 19\beta_3}} \Lambda}{4\pi r_h}
\end{equation}
Figure~\eqref{f8} displays the evolution of the metric function \( \psi(r) \) for a black hole surrounded by a quintessence field within the framework of \( f(R, \mathcal{L}_m, T) \) gravity, assuming a fixed cosmological constant \( \Lambda = 0.001 \). The curves are plotted for selected values of the parameter \( \beta_3 \), which are chosen such that the strong energy condition (SEC) remains satisfied, as discussed in Subsection~A. In contrast to the dust and radiation scenarios, where the spacetime tends toward the Schwarzschild solution at large distances, the presence of quintessence leads to significant departures in \( \psi(r) \) at large \( r \), highlighting the influence of the surrounding field on the asymptotic structure of the geometry.
\par
Equation~\eqref{e33} provides the expression for the Hawking temperature of a black hole surrounded by a quintessence field within the framework of \( f(R, \mathcal{L}_m, T) \) gravity. This relation captures the thermodynamic response of the system by incorporating the effects of both the horizon radius \( r_h \) ,$\Lambda$ and the matter-curvature coupling parameter \( \beta_3 \). The resulting expression defines a family of temperature profiles that reflect the influence of non-minimal coupling on the black hole's thermal characteristics.
Figure~\eqref{f9} depicts the variation of the Hawking temperature \( \textbf{T} \) as a function of the event horizon radius \( r_h \) for several representative values of \( \beta_3 \). As evident from the figure, the temperature remains strictly positive for all values considered, confirming the physical viability of the solutions. Additionally, a monotonic increase in temperature with decreasing \( r_h \) is observed. This behavior underscores the role of the coupling parameter \( \beta_3 \) in modulating the thermodynamics of the black hole, while still preserving the essential physical requirement of non-negative temperature.

\begin{figure}[H]
    \centering
    \includegraphics[width=1\linewidth]{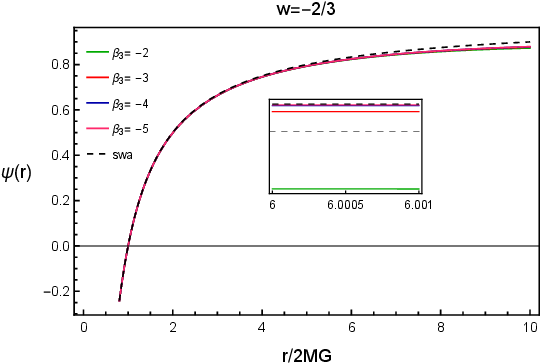}
    \caption{The behaviour of \( \psi(r) \) is analysed for different values of \( \beta_3 \), assuming $w=-2/3$ and a fixed cosmological constant \( \Lambda = 0.001 \).}
    \label{f8}
    \end{figure}
    \begin{figure}[H]
    \centering
    \includegraphics[width=1\linewidth]{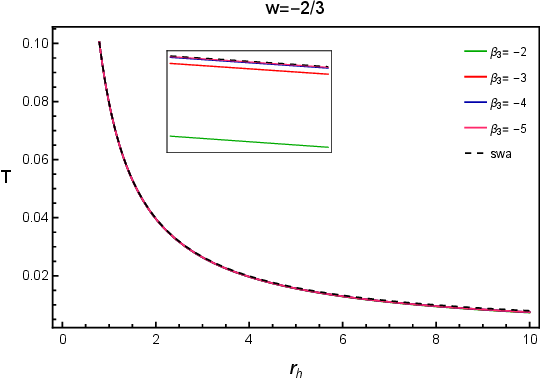}
    \caption{Equation \eqref{e33} is graphically represented as a function of the horizon radius \( r_h \) and the coupling parameter \( \beta_3 \), for a fixed cosmological constant \( \Lambda = 0.001 \) and an equation of state parameter \( w = -2/3 \) corresponding to a quintessence fluid. }
    \label{f9}
    \end{figure} 
\section{Black Hole solution for Non Liner function of $f(R,\mathcal L_{m},T)$ }\label{Sec4}
As a second case within the framework of \( f(R, \mathcal{L}_m, T) \) modified gravity, we consider a nonlinear functional form of the action given by
$f(R, \mathcal{L}_m, T) = \beta_1 R + \beta_2 \mathcal{L}_m T$,
where \( \beta_1 \) and \( \beta_2 \) are constant coupling parameters. This particular choice introduces direct coupling between the matter Lagrangian \( \mathcal{L}_m \) and the trace of the energy-momentum tensor \( T \), leading to nontrivial modifications of the gravitational dynamics. The corresponding gravitational field equations, obtained from the general expression in Eq.~\eqref{e9}, are derived straightforwardly and take the following form:
\begin{equation}\label{e34}
\begin{split}
\beta_1G_{\mu \nu}-\frac{1}{2}(\beta_2T\mathcal L_m)g_{\mu\nu}+\frac{1}{2} \mathcal L_m(\beta_2 T+2\beta_2 \mathcal L_m)g_{\mu \nu}\\
    =\Big(k+\frac{1}{2}(\beta_2 T+2\beta_2 \mathcal L_m)\Big)T_{\mu \nu}        
\end{split}
\end{equation}
To solve the modified field equations \eqref{e34} derived by inserting the nonlinear form of \( f(R, \mathcal{L}_m, T) \), we employ the same method outlined in Sec.\eqref{Sec3}. This approach leads to the following set of differential equations:
\begin{equation}\label{e35}
    -\frac{\beta_1}{r^2} + k \rho + \frac{1}{2}(-1 + w + 2w^2)\beta_2 \rho^2 + \frac{\beta_1 (\psi(r) + r \psi'(r))}{r^2} = 0
\end{equation}
\begin{equation}\label{e36}
  \frac{\beta_1 \psi'(r)}{r} 
+ \frac{1}{4} \left( 
\rho \left( -2 k_1 (1 + 3w) + (1 + w)^2 \beta_2 \rho \right) 
+ 2\beta_1 \psi''(r) 
\right) = 0  
\end{equation}
Consequently, the problem reduces to determining two unknown functions, \( \psi(r) \) and \( \rho(r) \), which can be found analytically by solving the two independent field equations. Eliminating \( \rho(r) \) from Eqs.\eqref{e35} and \eqref{e36} yields a differential equation solely in terms of the metric function \( \psi(r) \).
\begin{widetext}
\begin{equation} \label{e37} 
\begin{split}
&\frac{6 k^2 w^2 r^2 
+ 6 k w^2 r^2 \sqrt{
k^2 - \frac{2(-1 + w + 2w^2) \beta_1 \beta_2 (-1 + \psi(r) + r \psi'(r))}{r^2}}}{(1 + w)(-1 + 2w) \beta_2 r
} \\
&+\frac{\quad - (-1 + w + 2w^2) \beta_1 \beta_2 \left(
-1 - w + (1 + w)\psi(r) - 3(-1 + w) r \psi'(r) + (1 - 2w) r^2 \psi''(r)
\right)}{
(1 + w)(-1 + 2w) \beta_2 r
} = 0
\end{split}
\end{equation}
\end{widetext}
Solving the differential equation \eqref{e37} yields the black hole metric in modified \( f(R, \mathcal{L}_m, T) \) gravity in the presence of an anisotropic fluid.
However, due to the complexity of the differential equation, a direct analytical solution is not feasible. To overcome this challenge, we give an alternative approach, which allows us to systematically obtain analytical solutions. This method not only solves the problem but also provides valuable insight into the behavior of the metric function \( \psi(r) \) in the modified \( f(R, \mathcal{L}_m, T) \) gravity framework.
\par
\vspace{0.1cm}
(i) To solve the differential equation, we begin by imposing the condition that the quantity inside the square root in Eq.~\eqref{e37} is non-negative. Specifically, we set the term under the square root equal to \(\left(\frac{j}{r}\right)^2\) \eqref{e38}, where \(j\) is a constant parameter. This substitution guarantees that the expression remains real-valued and physically meaningful.

\begin{equation}\label{e38}  
k^2 - \frac{
2(-1 + w + 2w^2)\, \beta_1 \beta_2 \left(-1 + \psi(r) + r \psi'(r)\right)
}{r^2}
= \left( \frac{j}{r} \right)^2
\end{equation}
(ii) We observe that Eq.~\eqref{e38} is a first-order differential equation in the metric function \( \psi(r) \). By solving this equation, we obtain an explicit expression for \( \psi(r) \), where \( C_1 \) appears as an integration constant, which can be interpreted as being related to the black hole mass.
\begin{equation}\label{e39}
 \psi(r)=\frac{
    -j^2 r + \frac{k^2 r^3}{3}
}{  2r (-1 + w + 2w^2)\, \beta_1 \beta_2}
+ 1 + \frac{C_1}{r}
\end{equation}
(iii) The final step is to determine the value of the parameter \( j \). To do this, we consider Eq.~\eqref{e39} as a particular solution to the original differential equation \eqref{e37}. By substituting the obtained expression for \( \psi(r) \) from Eq.~\eqref{e39} back into Eq.~\eqref{e37}, we solve the resulting equation \eqref{e40}to determine the value of \( j \) \eqref{e41}.
\begin{equation}\label{e40}
\frac{
-j^2 (1 + w) + 
3 k r^2 \left( 4 \sqrt{\frac{j^2}{r^2}} w^2 - k (-1 + w + 4 w^2) \right)
}{
4 (1 + w) (r - 2r w)^2 \beta_2
}=0
\end{equation}
\begin{equation}\label{e41}
\begin{split}
 &   j = \frac{
6 k r w^2 - 
\sqrt{3} \sqrt{
k^2 r^2 \left( 1 + w^2 (1 + 2w)(-5 + 6w) \right)
}
}{
1 + w
}\\
&
j =\frac{
6 k r w^2 - 
\sqrt{3} \sqrt{
k^2 r^2 \left( 1 + w^2 (1 + 2w)(-5 + 6w) \right)
}
}{1 + w}
\end{split}
\end{equation}
\begin{equation}\label{e42}
   \psi(r)= 1 + 
\frac{
\frac{k^2 r^3}{3} - 
\frac{
r \left( 6 k r w^2 + \sqrt{3} \sqrt{
k^2 r^2 \left( 1 + w^2 (1 + 2w)(-5 + 6w) \right)
} \right)^2
}{
(1 + w)^2}
}{
2 r (-1 + w + 2w^2) \beta_1 \beta_2
}
+ \frac{C_1}{r}
\end{equation}
Equation \eqref{e42} represents the black hole metric solution in the modified \( f(R, \mathcal{L}_m, T) \) gravity theory in the presence of a surrounding fluid.By substituting the metric function obtained in Eq.~\eqref{e42} into Eq.~\eqref{e36}, we derive an equation \eqref{e44} involving the energy density \( \rho(r) \). Solving this equation \eqref{e44}, we obtain the explicit form of the energy density \( \rho(r) \) within the framework of modified \( f(R, \mathcal{L}_m, T) \) gravity in the presence of a surrounding fluid.
Where $C_1=-2M$
\begin{widetext}
\begin{equation}\label{e44}
\begin{split}
-\frac{18 \sqrt{3} k w^2 \sqrt{k^2 r^2 \left( 1 + w^2 \left( 1 + 2 w \right) \left( -5 + 6 w \right) \right)}}{r (1 + w)^3 (-1 + 2 w) \beta_2} + k \rho + \frac{1}{2} (-1 + w + 2 w^2) \beta_2 \rho^2 = \frac{k^2 \left( 4 + w \left( -1 + w \left( -23 + 18 w \left( -1 + 6 w \right) \right) \right) \right)}{(1 + w)^3 (-1 + 2 w) \beta_2}
\end{split}
\end{equation}
\end{widetext}
\subsection{Energy condition,mass and temparature}
In this subsection, we conduct a comprehensive analysis of the black hole metric structure and investigate the thermodynamic properties, including the black hole mass and Hawking temperature, for spacetimes surrounded by different matter fields: dust (\( w = 0 \)), radiation (\( w = 1/3 \)), and quintessence (\( w = -2/3 \)), within the framework of a non-linear functional form of \( f(R, \mathcal{L}_m, T) \) gravity. The corresponding expressions for the metric function, black hole mass, and Hawking temperature are given by Eqs.~\eqref{e44}, \eqref{e45}, and \eqref{e46}, respectively. The methodology adopted in this analysis parallels that discussed in detail in Subsections~A and B, where the linear case of \( f(R, \mathcal{L}_m, T) \) was examined.
\begin{widetext}
\begin{equation}\label{e44}
\begin{split}
ds^2 &=-\Big(1 + \frac{\frac{k^2 r^3}{3} - \frac{r \left( 6 k r w^2 + \sqrt{3} \sqrt{k^2 r^2 \left( 1 + w^2 \left( 1 + 2 w \right) \left( -5 + 6 w \right) \right)} \right)^2}{(1 + w)^2}}{2 r (-1 + w + 2 w^2) \beta_1 \beta_2} - \frac{2M}{r}\Big) \, dt^2 \\
&+ \frac{dr^2}{\Big(1 + \frac{\frac{k^2 r^3}{3} - \frac{r \left( 6 k r w^2 + \sqrt{3} \sqrt{k^2 r^2 \left( 1 + w^2 \left( 1 + 2 w \right) \left( -5 + 6 w \right) \right)} \right)^2}{(1 + w)^2}}{2 r (-1 + w + 2 w^2) \beta_1 \beta_2} - \frac{2M}{r}\Big)} + r^2 \left( d\theta^2 + \sin^2\theta \, d\phi^2 \right)
\end{split}
\end{equation}
\begin{equation}\label{e45}
M(r_h)=-\frac{1}{2} r_h \left( -1 - \frac{\frac{k^2 r_h^3}{3} - \frac{r_h \left( 6 k r_h w^2 + \sqrt{3} \sqrt{k^2 r_h^2 \left( 1 + w^2 \left( 1 + 2 w \right) \left( -5 + 6 w \right) \right)} \right)^2}{(1 + w)^2}}{2 r_h (-1 + w + 2 w^2) \beta_1 \beta_2} \right)
\end{equation}
\begin{equation}\label{e46}
\small{\textbf{T}=\frac{k^2 r_h^2 \left( -4 + w + 23 w^2 + 18 w^3 - 108 w^4 \right) - 18 \sqrt{3} k r_h w^2 \sqrt{k^2 r_h^2 
\left( 1 + w^2 \left( 1 + 2 w \right) \left( -5 + 6 w \right) \right)} + (1 + w)^3 (-1 + 2 w) \beta_1 \beta_2}{4 \pi r_h (1 + w)^3 (-1 + 2 w) \beta_1 \beta_2}} 
\end{equation}   
\end{widetext}
To investigate the physical viability of the anisotropic matter distribution considered in this framework, we now examine the conditions under which the \textit{Strong Energy Condition (SEC)} \eqref{e22} is satisfied. In the context of modified gravity theories such as \( f(R, \mathcal{L}_m, T) \), the SEC plays a crucial role in constraining the parameter space of the model. By applying the derived expressions for the energy density \( \rho (r)\), radial pressure \( p_r \), and tangential pressure \( p_t \), we determine the regions in the parameter space (e.g., involving \( \beta_1 \), and \( \beta_2 \)) where the SEC holds. This analysis provides essential insight into the nature of the effective matter content and the corresponding gravitational dynamics.
\par
Now, we consider specific cases by fixing the equation of state parameter \( w \), corresponding to different surrounding matter fields. For each fixed value of \( w \), we analyse the conditions under which the Strong Energy Condition (SEC) is satisfied. These conditions impose constraints on the coupling parameters, such as \( \beta_1 \) and  \( \beta_2 \), ensuring the physical viability of the matter distribution. Once the SEC-satisfying parameter ranges are identified, we proceed to investigate the corresponding black hole solutions by analysing the behaviour of the metric function \( \psi(r) \) and the associated Hawking temperature. This approach allows us to systematically explore how the nature of the surrounding matter influences the spacetime geometry and thermodynamic properties of black holes in \( f(R, \mathcal{L}_m, T) \) gravity. In addition,  we assume \( \beta_1 = 1 \) and  the gravitational constant \( k = 1 \) throughout the analysis.

 \subsubsection{Black hole surrounded by a dust field}
By substituting the equation of state parameter \( w = 0 \), which corresponds to a dust-like matter distribution, into Eq.~\eqref{e22} for the non-linear case of \( f(R, \mathcal{L}_m, T) \) gravity, we obtain the condition for the Strong Energy Condition (SEC). The SEC condition in this case is given by:
 \begin{equation}\label{e47}
 \rho + p_r +  2p_t=\frac{4}{\beta_2}\geq0
 \end{equation}
 From Eq.~\eqref{e47}, it is evident that the strong energy condition (SEC) is satisfied only when the parameter \( \beta_2 \) takes positive values, i.e., \( \beta_2 > 0 \). The corresponding behaviour of the SEC for Black hole surrounded by a dust field can also be visualised in Figure~\eqref{f10}.
 \par
 After identifying the conditions under which the strong energy condition (SEC) is satisfied, we substitute the corresponding values into Eqs.~\eqref{e44} and \eqref{e46} to derive the black hole metric and the associated Hawking temperature for the case where \( f(R, \mathcal{L}_m, T) \) is described by a nonlinear function within the framework of modified \( f(R, \mathcal{L}_m, T) \) gravity.
 \par
 Figure~\eqref{f11} illustrates the behavior of the black hole metric function \( \psi(r) \) for a black hole surrounded by a dust field (\( w = 0 \)), in the framework of modified \( f(R, \mathcal{L}_m, T) \) gravity, where the function \( f(R, \mathcal{L}_m, T) \) is taken to be nonlinear. The figure demonstrates the influence of the parameter \( \beta_2 \) on the behaviour of the metric function \( \psi(r) \) within the context of \( f(R, \mathcal{L}_m, T) \) gravity with a nonlinear functional form. As \( r \) increases, the curves corresponding to different values of \( \beta_2 \) increasingly diverge from one another. Notably, the extent of divergence becomes less pronounced with higher values of \( \beta_2 \), indicating that larger couplings lead to a smoother radial evolution of the metric function. This behaviour contrasts markedly with the linear case of \( f(R, \mathcal{L}_m, T) \) gravity, as well as with predictions from General Relativity, highlighting the distinct imprints of nonlinear matter-curvature couplings on the spacetime structure.
 \par
 Figure~\eqref{f12} presents the variation of the Hawking temperature \( \textbf{T} \) as a function of the event horizon radius \( r_h \) for a black hole surrounded by a dust field (\( w = 0 \)) within the framework of modified \( f(R, \mathcal{L}_m, T) \) gravity. In this analysis, the function \( f(R, \mathcal{L}_m, T) \) is assumed to be nonlinear, incorporating a nontrivial coupling between matter and geometry. The temperature profile reveals several notable features. Firstly, the Hawking temperature remains strictly positive across the entire range of horizon radius considered, ensuring the physical validity of the solutions. A distinct non-monotonic behaviour is observed: the temperature initially increases as the horizon radius decreases, and then decreases with increasing \( r_h \). Interestingly, beyond a certain threshold, the temperature begins to rise again with increasing horizon radius. This nontrivial thermal behaviour is directly influenced by the nonlinear matter-curvature coupling encoded in the parameter \( \beta_2 \). As \( \beta_2 \) increases, the rate of change of temperature with respect to \( r_h \) becomes more gradual, indicating that the coupling softens the thermal response of the black hole system. This complex thermodynamic behaviour sharply contrasts with that found in both the linear \( f(R, \mathcal{L}_m, T) \) models and General Relativity, where the temperature typically decreases monotonically with increasing horizon radius. The presence of a turning point in the temperature profile highlights the rich and distinctive thermodynamic structure introduced by nonlinear modifications to the gravitational theory.

\begin{figure}[H]
    \centering
    \includegraphics[width=1\linewidth]{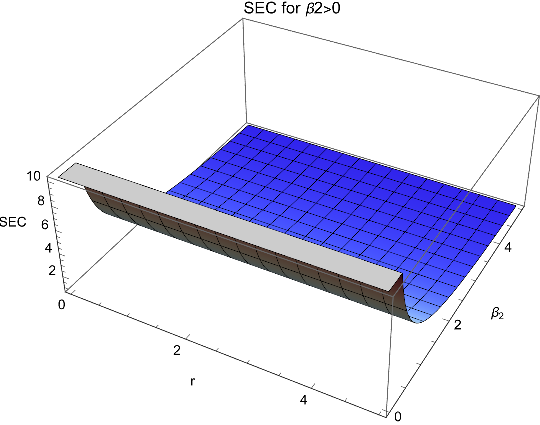}
    \caption{}
    \label{f10}
    \end{figure}
    \begin{figure}[H]
    \centering
    \includegraphics[width=1\linewidth]{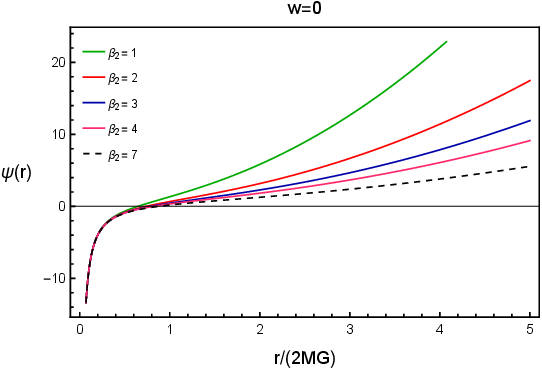}
    \caption{}
    \label{f11}
    \end{figure} 
    \begin{figure}[H]
    \centering
    \includegraphics[width=1\linewidth]{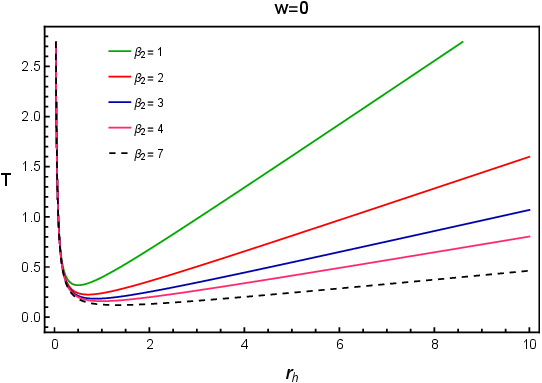}
    \caption{}
    \label{f12}
    \end{figure}
 \subsubsection{Black hole surrounded by a quintessence  Field}
Now, in the second case, we consider a black hole surrounded by a quintessence field, characterised by the equation of state parameter \( w = -\frac{2}{3} \). Substituting this value into Eq.~\eqref{e22}, and following the procedure outlined in the preceding subsection, we derive the condition under which the strong energy condition (SEC) is satisfied in the context of the nonlinear modification of \( f(R, \mathcal{L}_m, T) \) gravity. The resulting form of the SEC for this configuration is given by the following expression:
 \begin{equation}\label{e48}
 \rho + p_r +  2p_t=\frac{-9 + 9 \sqrt{381 + \frac{144 \sqrt{7} \, r}{\sqrt{r^2}}}}{7 \beta_2}
\geq0
 \end{equation}
 From Eq.~\eqref{e48}, it is evident that the strong energy condition (SEC) is satisfied when \( r > 0 \) and \( \beta_2 > 0 \). This indicates that the energy-momentum tensor respects the SEC throughout the spacetime exterior to the black hole horizon for positive values of the coupling parameter \( \beta_2 \). The corresponding behaviour of the SEC for a black hole surrounded by a quintessence field is illustrated in Figure~\eqref{f13}, where the parameter space ensuring SEC validity is clearly visualised.
\par
Figures~\eqref{e14} and \eqref{e15} depict the behaviour of the metric function and the corresponding Hawking temperature for a black hole surrounded by a quintessence field, within the framework of the modified \( f(R, \mathcal{L}_m, T) \) gravity.We observe that the metric function \( \psi(r) \) exhibits divergent behavior as the radial coordinate \( r \) increases. Furthermore, an increase in the coupling parameter \( \beta_2 \) leads to a noticeable reduction in the slope of the metric function, indicating that the divergence becomes less steep. This behavior highlights the significant role of the nonlinear matter-geometry coupling in shaping the spacetime geometry.
We observe from Fig.~\eqref{f15} that the Hawking temperature remains strictly positive for all values of the horizon radius \( r_h \). The temperature initially decreases with increasing \( r_h \), reaches a minimum, and subsequently begins to increase. This non-monotonic behavior reflects the intricate influence of the matter-geometry coupling on the thermodynamic properties of the black hole.
 \begin{figure}[H]
    \centering
    \includegraphics[width=1\linewidth]{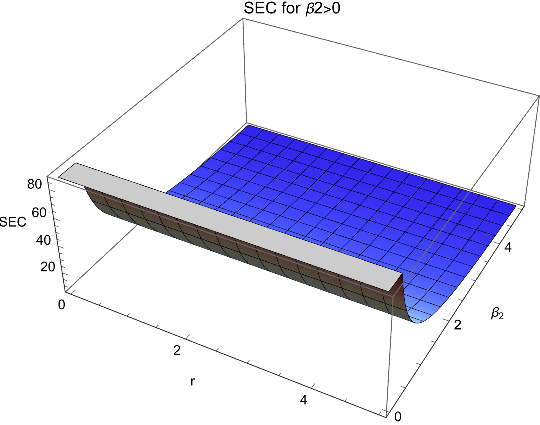}
    \caption{}
    \label{f13}
    \end{figure}
    \begin{figure}[H]
    \centering
    \includegraphics[width=1\linewidth]{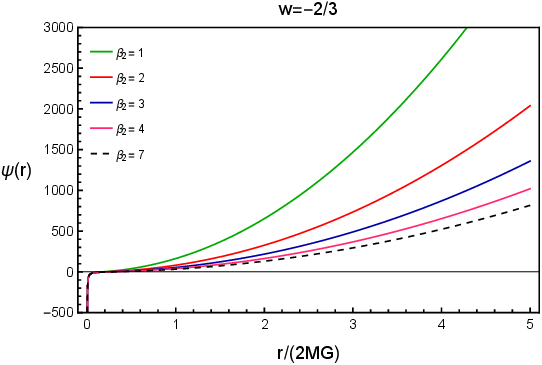}
    \caption{}
    \label{f14}
    \end{figure} 
    \begin{figure}[H]
    \centering
    \includegraphics[width=1\linewidth]{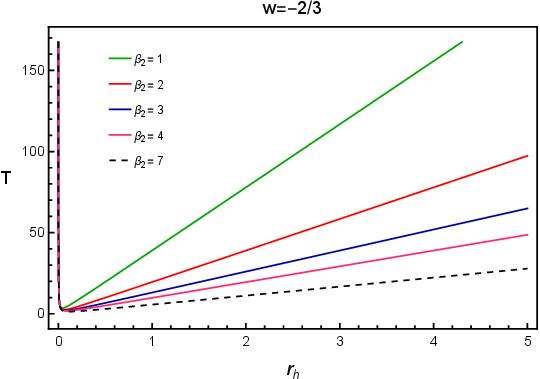}
    \caption{}
    \label{f15}
    \end{figure}
    \subsubsection{Black hole surrounded by a radiation field}
    In the third case, we consider a black hole surrounded by a radiation field, described by the equation of state parameter \( w = \frac{1}{3} \). Substituting this value into Eq.~\eqref{e22} and applying the same procedure outlined in the previous subsection, we derive the specific conditions under which the strong energy condition (SEC) is satisfied in the context of a nonlinear formulation of \( f(R, \mathcal{L}_m, T) \) gravity. The corresponding expression for the SEC, applicable to this radiation-dominated configuration, is given by:
     \begin{equation}\label{e49}
 \rho + p_r +  2p_t=\frac{18 + 9 \sqrt{12 + \dfrac{6 \sqrt{3} r}{\sqrt{r^2}}}}{4 \beta_2}\geq0
 \end{equation}
 Equation~\eqref{e49} reveals that the strong energy condition (SEC) is upheld when \( r > 0 \) and \( \beta_2 > 0 \). This condition ensures that the energy-momentum tensor remains physically consistent across the exterior region of the black hole spacetime, given a positive matter-curvature coupling parameter \( \beta_2 \). Figure~\eqref{f16}visualizes the corresponding SEC behavior for a black hole immersed in a radiation field, clearly identifying the region of parameter space where the SEC remains valid.
 \par
 Figures~\eqref{f17} and \eqref{f18} present the variation of the metric function \( \psi(r) \) and the Hawking temperature for a black hole enveloped by a radiation field in the context of modified \( f(R, \mathcal{L}_m, T) \) gravity. The metric function \( \psi(r) \) shows an increasing divergence as the radial coordinate \( r \) grows. Additionally, an increase in the coupling parameter \( \beta_2 \) leads to a reduction in the steepness of this divergence, suggesting a moderating effect of the nonlinear matter-geometry coupling on the spacetime curvature.

As illustrated in Fig.~\eqref{f17}, the Hawking temperature remains positive across all values of the horizon radius \( r_h \). The temperature profile is non-monotonic, initially decreasing with \( r_h \), reaching a minimum, and subsequently rising again. This behavior highlights the complex influence of matter-geometry interactions on the thermodynamic properties of the black hole. 
 \begin{figure}[H]
    \centering
    \includegraphics[width=1\linewidth]{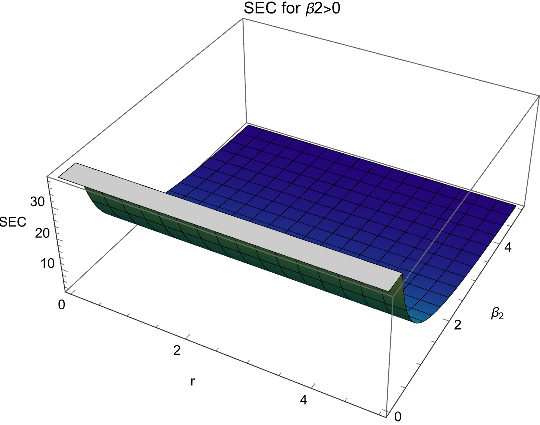}
    \caption{}
    \label{f16}
    \end{figure}
    \begin{figure}[H]
    \centering
    \includegraphics[width=1\linewidth]{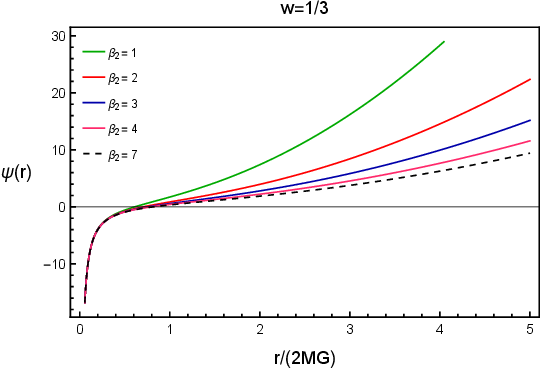}
    \caption{}
    \label{f17}
    \end{figure} 
    \begin{figure}[H]
    \centering
    \includegraphics[width=1\linewidth]{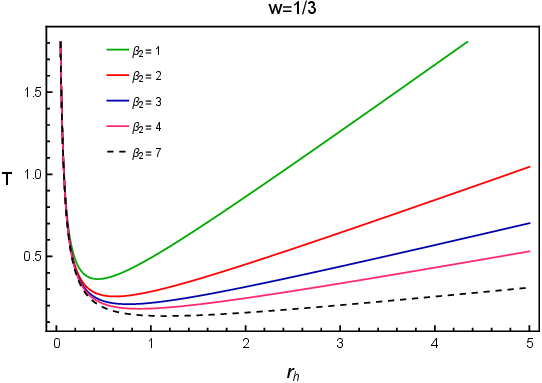}
    \caption{}
    \label{f18}
    \end{figure}
    \section{SUMMARY AND CONCLUSIONS}\label{sec4}
    In this study, we construct a new black hole solutions within the framework of modified \( f(R, \mathcal{L}_m, T) \) gravity, where the black hole is surrounded by an anisotropic fluid. We consider both linear and nonlinear forms of   \( f(R, \mathcal{L}_m, T) \) in order to comprehensively analyze the effects of matter-geometry coupling on the resulting spacetime geometry and thermodynamic behavior. The strong energy condition (SEC) is imposed to ensure physically viable matter distributions, and we determine the parameter regimes under which the SEC is satisfied. Specifically, we investigate three distinct matter field configurations characterized by the equation of state parameter \( w = 0, 1/3, -2/3 \), corresponding to dust, radiation, and quintessence fields, respectively. For each of these cases, we derive the corresponding black hole metric and examine the thermodynamic properties, including the black hole mass and Hawking temperature. Our analysis reveals how different forms of the coupling function and surrounding matter content influence the structure and thermal behavior of black holes in this modified theory of gravity.This study presents, for the first time, an explicit solution to the gravitational field equations in the framework of \( f(R, \mathcal{L}_m, T) \) gravity for a black hole surrounded by an anisotropic fluid, thereby contributing a novel result to the existing literature on modified gravity theories.
    \par
   (i) We begin our analysis by deriving the field equations from the action \eqref{e1} for the modified \( f(R, \mathcal{L}_m, T) \) gravity theory. As part of this derivation, we also determine the non-conservation condition for the energy-momentum tensor \( T_{\mu\nu} \), which arises naturally in this modified framework. We then consider a linear functional form for \( f(R, \mathcal{L}_m, T) \), and adopt a static, spherically symmetric spacetime characterized by a single metric function \( \psi(r) \). The black hole is assumed to be surrounded by an anisotropic fluid distribution.
 By solving two independent components of the modified field equations under these assumptions, we obtain an exact analytical form of the black hole metric function \( \psi(r) \) corresponding to the linear case of \( f(R, \mathcal{L}_m, T) \). After obtaining the solution, we derive the condition under which the strong energy condition (SEC) is satisfied, which imposes constraints on the matter-curvature coupling parameters.
Utilizing this condition, we then explore the behavior of the metric function and the Hawking temperature for three specific cases of the state parameter \( w \), namely \( w = 0 \), \( w = \frac{1}{3} \), and \( w = -\frac{2}{3} \), corresponding to dust, radiation, and quintessence field configurations, respectively. This analysis reveals how different surrounding matter fields influence the spacetime structure and thermodynamic properties of the black hole in the presence of matter-geometry coupling.
\par
(ii) In the next stage of our analysis, we consider a nonlinear form of the function \( f(R, \mathcal{L}_m, T) \) to investigate how nonlinearity affects the gravitational field. In this scenario, the modified field equations become more intricate, involving two unknown functions: the energy density \( \rho(r) \) and the metric function \( \psi(r) \), which must be determined simultaneously. Unlike the linear case, these equations cannot be solved directly.To address this challenge, we propose a systematic approach to solve the differential equations and obtain an analytic expression for the metric function \( \psi(r) \), which describes the spacetime structure of a black hole surrounded by an anisotropic fluid within the nonlinear \( f(R, \mathcal{L}_m, T) \) gravity framework. After deriving the metric solution, we analyze the conditions under which the strong energy condition (SEC) holds, thereby placing constraints on the matter-curvature coupling parameters. Using these constraints, we then explore the behavior of both the metric function and the Hawking temperature for three specific cases of the equation of state parameter \( w \): dust (\( w = 0 \)), radiation (\( w = \frac{1}{3} \)), and quintessence (\( w = -\frac{2}{3} \)).
\par
(iii) After examining the metric function and Hawking temperature in both the linear and nonlinear formulations of \( f(R, \mathcal{L}_m, T) \) gravity, we observed significant differences in their behaviour. In the linear case, the metric function \( \psi(r) \) exhibits standard behaviour, asymptotically converging to the Schwarzschild solution as \( r \) increases. Similarly, the Hawking temperature increases for small horizon radius and gradually decreases as the horizon radius grows. In contrast, the nonlinear case reveals distinct and nontrivial phenomena. The metric function and Hawking temperature become highly sensitive to the matter-geometry coupling parameter \( \beta_2 \). Notably, the metric function \( \psi(r) \) displays a divergent trend from each other with increasing \( r \), deviating from the Schwarzschild behaviour observed in the linear case. However, increasing \( \beta_2 \) tends to soften this divergence, indicating a damping effect induced by stronger coupling. Additionally, the Hawking temperature in the nonlinear case does not monotonically decrease. Instead, it initially decreases with increasing horizon radius \( r_h \), reaches a minimum, and then begins to increase. This non-monotonic trend becomes more aligned with the linear case as \( \beta_2 \) increases, highlighting a smooth transition between the two regimes. This behaviour underscores the rich and complex nature of matter-geometry interactions in the nonlinear framework and represents a particularly intriguing outcome of our study.
\par
(iv)In the linear case, the black hole exhibits features and thermodynamic behaviour consistent with those observed in Einstein and Lovelock gravity. However, this correspondence does not hold in the nonlinear formulation of \( f(R, \mathcal{L}_m, T) \) gravity. It is therefore compelling to investigate how different functional forms of \( f(R, \mathcal{L}_m, T) \) influence the spacetime geometry and thermodynamic characteristics of black holes.
Furthermore, incorporating additional matter sources—such as a Maxwell field—could enrich the thermodynamic structure and offer deeper insights into the interplay between nonlinearity and matter-geometry coupling in this modified gravity framework.
\section{Acknowledgements}
AG is thankful to IIEST, Shibpur, India, for providing an Institute
Fellowship (JRF).
\bibliographystyle{naturemag}
\bibliography{bibliography}
\end{document}